\documentclass[journal]{IEEEtran}
\usepackage{amsmath,amsfonts}
\usepackage{array}
\usepackage[caption=false,font=normalsize,labelfont=sf,textfont=sf]{subfig}
\usepackage{textcomp}
\usepackage{stfloats}
\usepackage{url}
\usepackage{verbatim}
\usepackage{graphicx}
\usepackage{cite}
\usepackage{braket}
\usepackage{amssymb}
\usepackage{yhmath}
\usepackage{color}
\usepackage{orcidlink}
\usepackage{geometry}
\usepackage{setspace}
\usepackage{orcidlink}
\usepackage[linesnumbered,ruled,commentsnumbered,longend]{algorithm2e}

\usepackage{hyperref}
\hypersetup{hypertex=true,
	colorlinks=true,
	linkcolor=blue,
	anchorcolor=red,citecolor=blue}
\allowdisplaybreaks[2] 

\begin{document}
\title{Six-Dimensional Movable Antenna Enabled Wideband THz Communications}
\author{Wencai Yan, \emph{Member, IEEE}, Wanming Hao, \emph{Senior Member, IEEE}, Yajun Fan, Yabo Guo, Qingqing Wu, \emph{Senior Member, IEEE}, Xingwang Li, \emph{Senior Member, IEEE}
\thanks{W. Yan, Y. Fan and Y. Guo are with the Key Laboratory of Grain Information Processing and Control (Henan University of Technology), Ministry of Education, Zhengzhou 450001, China, and also with the College of Information Science and Engineering, Henan University of Technology, Zhengzhou 450001, China. (E-mail: {yanwencai, yjfan, ybguo}@haut.edu.cn)}
\thanks{W. Hao is with the School of Electrical and Information Engineering, Zhengzhou University, Zhengzhou 450001, China. (E-mail: iewmhao@zzu.edu.cn)}
\thanks{Q. Wu is with the Department of Electronic Engineering, Shanghai Jiao Tong University, Shanghai 200240, China (E-mail: qingqingwu@sjtu.edu.cn)}
\thanks{X. Li is with the School of Physics and Electronic Information Engineering, Henan Polytechnic University, Jiaozuo 454000, China, and X. Li is also  with the National Mobile Communications Research Laboratory, Southeast University, Nanjing 214135, China (email: lixingwang@hpu.edu.cn)}

}

\maketitle

\begin{abstract}
In this paper, we investigate a six-dimensional movable antenna (6DMA)-enabled wideband terahertz (THz) communication system with sub-connected hybrid beamforming architecture at the base station (BS). In particular, the three-dimensional (3D) position and 3D rotation of each 6DMA surface can be flexibly reconfigured to mitigate the beam squint effects instead of introducing costly true-time-delay devices. We first analyze the normalized array gain in the 6DMA-enabled wideband THz systems based on the beam squint effects. Then, we formulate a sum-rate maximization problem via jointly optimizing 3D positions, 3D rotations, and hybrid analog/digital beamforming. To solve the non-convex problem, an alternating optimization algorithm is developed that decomposes the original problem into three subproblems, which are solved alternately. Specifically, given the positions and rotations of 6DMA surfaces, we first reformulate the objective function and design a semidefinite relaxation-based alternating minimization scheme to  obtain the hybrid analog/digital beamforming. 
Then, the positions and rotations of the 6DMA surfaces are further optimized through a feasible gradient descent procedure. The final solutions are obtained by repeating the above procedure until convergence. Numerical results demonstrate the superior performance of the proposed scheme compared  with conventional fixed-position antenna architectures.
\end{abstract}


\IEEEpeerreviewmaketitle

\section{Introduction}
To meet the requirement of high speed wireless date  transmission, high-frequency bands such as terahertz (THz) are considered promising candidates for providing abundant frequency resources in future sixth-generation (6G) communications~\cite{refI_0}-\cite{refI_3}. However, THz signals are severely affected by free-space loss and non-line-of-sight (NLoS) attenuation. Consequently, massive multiple-input multiple-output (MIMO) architectures have been recognized as an energy-efficient solution to generate high-gain directional beams and effectively expand THz communication coverage~\cite{refI_4}, \cite{refI_4_1}, \cite{refI_4_2}. Moreover, to enable practical deployment of large-scale antenna arrays, hybrid beamforming has developed as a pivotal technique that connects a large number of antennas to a small number of RF chains through phase shifters (PSs)~\cite{refI_5}, \cite{refI_5_1}, \cite{refI_5_2}. This configuration is typically realized by two primary architectures: fully-connected and sub-connected architecture~\cite{refI_6},~\cite{refI_7}.  

Although hybrid beamforming architectures perform well in narrowband communications, there exist beam squint effects in wideband THz communications, which degrades system performance. Ideally, the phase shifts required for constructive interference are frequency-dependent, whereas the phase shifts provided by PSs in hybrid beamforming architecture are frequency-independent ~\cite{refI_8}. This discrepancy causes beams at non-central frequencies to deviate from the intended directions, which reduces the beamforming gain~\cite{refI_9}, \cite{refI_9_1}, \cite{refI_9_2}.
Furthermore, most existing works addressing the beam squint effects are limited to fixed-position antenna schemes, which inherently lack the flexibility to adapt antenna configurations. To overcome this limitation, six-dimensional movable antenna (6DMA) has recently  emerged~\cite{refI_10}-\cite{refI_11_2}. Through jointly optimizing the three-dimensional (3D) positions and 3D rotations of all surfaces within a given deployment space, the 6DMA-equipped transceivers can dynamically assign antenna resources based on the users distribution to maximize the beamforming gain \cite{refI_12}. Therefore, investigating effective configuration strategies for 6DMA is crucial and useful to mitigate the beam squint effects in wideband THz communications.

\subsection{Related Works}
To further utilize the spatial DoFs, the authors of~\cite{refI_19} propose the 6DMA architecture and develop a achievable sum-rate maximization problem where 3D positions and rotations are jointly optimized under the practical movement constraints on 6DMA surfaces. In~\cite{refI_20}, a 6DMA-assisted base station (BS) is considered with discrete positions and rotations available. To overcome this limitation, a new online learning optimization method  is introduced that operates without prior knowledge of user channel distribution. Furthermore, the authors of~\cite{refI_21} propose a  low-complexity framework for 6DMA-aided communication systems with statistical channel information (SCI) estimation and SCI-based 6DMA positions and rotations optimization. Based on the estimated SCI, a sequential optimization strategy jointly considers 6DMA rotations and positions, achieving a high sum-rate with reduced complexity. In addition, a passive 6DMA-assisted multiuser uplink system has been studied, where an optimization problem is formulated to maximize the achievable sum-rate via jointly optimizing the surfaces positions, rotations, and reflection coefficients under practical movement constraints on passive 6DMA surfaces~\cite{refI_22}, \cite{refI_22_1}. However, most previous works on 6DMA assumed the fully-digital architecture, where each antenna is connected to a dedicated RF chain, resulting in high hardware complexity. Thus, the authors of~\cite{refI_23} investigate a 6DMA-assisted multi-user hybrid  beamforming technique. Meanwhile, by simultaneously considering the radiation patterns and polarization conditions of actual directional antennas, the performance of 6DMA systems is precisely characterized. Although the aforementioned studies have demonstrated the performance advantages of 6DMA, they do not consider the wideband THz communication scenarios.

While THz communications own abundant bandwidth, it also introduces significant challenges, particularly the beam squint effects.  Currently, relevant researches have explored several solutions to overcome the beam squint effects based on fixed-position antenna  architectures. One straightforward method involves substituting all PSs with true-time-delays (TTDs), which provide  frequency-proportional phase shifts by generating a constant delay across different subcarriers, and thus the beam squint effects can be effectively mitigated~\cite{refI_A}. However, TTDs consume higher
power consumption and hardware complexity than PSs, so it is impractical to directly replace all PSs with TTDs. 
A more feasible alternative integrates a limited number of TTDs between the RF chains and PSs, enabling the conventional one-dimensional analog beamforming transformed to two-dimensional analog beamforming by the joint control of PSs and TTDs~\cite{refI_13}, \cite{refI_14}, \cite{refI_15}.  In such configurations, TTDs are typically arranged in parallel, with each unit requiring independent configuration and a large time delay range, particularly for large arrays~\cite{refI_16}. Consequently, 
to enhance energy efficiency, a dynamic-subarray architecture with fixed TTDs has been developed~\cite{refI_17}.
Despite these performance gains, the improvement remains limited. To reduce the hardware complexity at the BS, the authors of~\cite{refI_18} present a double-layer TTD scheme, which overcomes the maximum delay compensation limitation inherent to conventional single-layer TTD schemes. However, the aforementioned approaches mainly depend on TTDs for phase compensation, which significantly increase hardware costs. 

Therefore, to alleviate the beam squint effects, several studies have explored antenna position optimization as a potential solution. For example, the authors of~\cite{refI_24} aim to maximize the minimum analog beamforming gain over the entire frequency subcarriers through appropriate adjustment of movable antenna positions. Similarly, ~\cite{refI_25} formulates an achievable sum-rate maximization problem by jointly optimizing BS beamforming, movable-element simultaneously transmitting and reflecting surface beamforming, and positions of movable elements. While above schemes can only partially alleviate the beam squint effects, the performance remains limited. In this paper, we introduce a 6DMA-based approach that effectively suppresses beam squint by jointly optimizing the 3D positions and 3D rotations of 6DMA surfaces.

\subsection{Main Contributions}
Inspired by the above advancements, we investigate the beam squnit effects and beamforming optimization problem in the 6DMA-enabled wideband THz communications, and the main contributions are summarized as~follows.

\begin{itemize}
	\item[$\bullet$]
	To address the beam squint effects with low hardware complexity, we introduce 6DMA to the wideband THz communication systems. The 6DMA provides a dynamic spatial compensation mechanism that mitigates beam squint by physically counteracting frequency-dependent phase shifts. We first derive the normalized array gain and analyze the beam squint effects. It demonstrates that through joint optimization of the 3D positions and rotations of 6DMA surfaces, beam squint effects in the wideband THz communication systems can be effectively suppressed without adding extra hardware. 
\end{itemize}
\begin{itemize}
	\item[$\bullet$]
	Moreover, a sub-connected hybrid beamforming architecture is employed at the BS to further reduce hardware complexity. Based on this structure, we formulate a sum-rate maximization problem that jointly optimizes 3D positions, 3D rotations of 6DMA surfaces and the hybrid analog/digital beamforming. Specifically, the optimization accounts for the unit-modulus constraint imposed by PSs, the transmit power constraint, and practical deployment and mobility constraints of 6DMAs. Such as, the minimum inter-spacing between any two 6DMA surfaces, as well as rotation constraints to prevent signal reflection and blockage caused by the BS central processing unit~(CPU).
\end{itemize}
\begin{itemize}
	\item[$\bullet$]
	To solve the non-convex optimization problem, we develop an alternating optimization (AO) algorithm that decomposes it into three subproblems and solves them alternately. Specifically, given the positions and rotations of all 6DMA surfaces, we first reformulate the objective function and devise a semidefinite relaxation (SDR)-based alternating minimization scheme to obtain the hybrid analog/digital beamforming. Then, the positions and rotations of the 6DMA surfaces are further optimized through a feasible gradient descent procedure. The final solution is obtained by iteratively repeating the above steps until convergence. Numerical results demonstrate the superior performance of 6DMA in wideband THz communications.
\end{itemize}

The remainder of this paper is organized as follows. Section II outlines the system model and analyzes the beam squint effects. Section III presents the problem formulation and the proposed  solution. Section IV evaluates the approach through numerical simulations, with final conclusions drawn in Section~V. 

Notations: Scalars, vectors, and matrices are represented by lower-case, boldface lower-case and boldface upper-case letters, respectively. $(\cdot)^{T}$ and $(\cdot)^{H}$ represent the transpose and Hermitian transpose, respectively.   $\lceil \cdot \rceil$ is the ceil function.  $\left\|\cdot \right\|$ means the Frobenius norm. $\mathbb{E}[\cdot]$ is the statistical expectation. $\operatorname{Tr}(\mathbf{A})$ and $\operatorname{vec}(\mathbf{A})$ indicate the trace and vectorization, respectively. $\mathbb{C}^{x \times y}$ denote the set of all $x \times y$ complex-valued matrices.  $\otimes$ means the Kronecker products between two matrices. 
\begin{figure}[t]
	\centering
	\includegraphics[height=1.9in,width=3.6in]{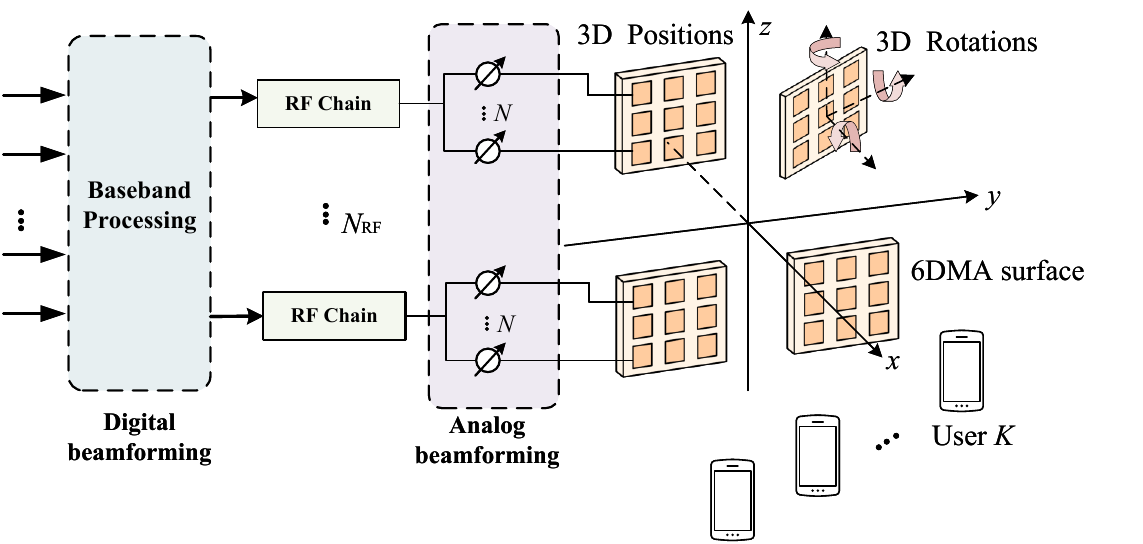}
	\caption{6DMA-enabled communication system with hybrid beamforming.}
	\label{fig_1}
\end{figure}

\section{System Model and Beam Squint Mitigation}
In this section, a 6DMA-enabled downlink multi-user THz communication system and the corresponding channel model are first described. Then, we analyze the beam squint effects.

\subsection{System Model}
As presented in Fig. \ref{fig_1}, we investigate a 6DMA-enabled downlink THz wideband communication system, where BS employs $S$ 6DMA surfaces to serve $K$ single-antenna users. Meanwhile, the BS adopts a sub-connected hybrid beamforming with $N_{\rm{RF}}$ radio frequency (RF) chains. There are $N$ antennas on each 6DMA surface, and thus the total number of antennas equipped at the BS is $N_t= S N$. Within each 6DMA surface, antennas are first individually connected to a dedicated PS, which are further connected to a corresponding RF chain. In particular, 6DMA surfaces connects to the CPU through retractable and rotatable rods equipped with flexible wires, enabling CPU to perform joint adjustment of 3D positions and 3D rotations~\cite{refII_1}.

The position and rotation of the $s$-th 6DMA surface are expressed as $\mathbf{p}_s=[x_s, y_s, z_s]^T \in \mathcal{D}$ and $\mathbf{u}_s=[\alpha_s, \beta_s, \gamma_s]^T  $, respectively, where $\mathcal{D}$ represents the movable/rotated 3D region. Here, $x_s$, $y_s$ and $z_s$ denote the coordinates of the $s$-th 6DMA’s center in the global Cartesian coordinate system (CCS) o-xyz, where the reference position of the 6DMA-enabled BS serves as the origin. $\alpha_s \in [0, 2 \pi)$, $\beta_s \in [0, 2 \pi)$ and $\gamma_s \in [0, 2 \pi)$ represent the rotation angles about the x-axis, y-axis and z-axis, respectively. Let $\mathbf{r}_{s,n}(\mathbf{p}_s, \mathbf{u}_s)$ denote the position of the $n$-th element of the $s$-th 6DMA surface within the global CCS, which is formulated as
\begin{equation}
\mathbf{r}_{s,n}(\mathbf{p}_s, \mathbf{u}_s)=\mathbf{p}_s+\mathbf{R}(\mathbf{u}_s) \bar{\mathbf{r}}_n,
\end{equation}
where $\bar{\mathbf{r}}_n$ denotes the position of the $n$-th antenna of the 6DMA surface in the local CCS. $\mathbf{R}(\mathbf{u}_s)$ represents the rotation matrix, which is defined as
\begin{equation}
	\begin{aligned}
	& \mathbf{R}\left(\mathbf{u}_s\right)=\\&\left[\begin{array}{ccc}
		c_{\beta_s} c_{\gamma_s} & c_{\beta_s} s_{\gamma_s} & -s_{\beta_s} \\
		s_{\beta_s} s_{\alpha_s} c_{\gamma_s}-c_{\alpha_s} s_{\gamma_s} & s_{\beta_s} s_{\alpha_s} s_{\gamma_s}+c_{\alpha_s} c_{\gamma_s} & c_{\beta_s} s_{\alpha_s} \\
		c_{\alpha_s} s_{\beta_s} c_{\gamma_s}+s_{\alpha_s} s_{\gamma_s} & c_{\alpha_s} s_{\beta_s} s_{\gamma_s}-s_{\alpha_s} c_{\gamma_s} & c_{\alpha_s} c_{\beta_s}
	\end{array}\right],
	\end{aligned}
\end{equation}
where $c_x = cos(x)$ and $s_x = sin(x)$~\cite{refII_2}.

The orthogonal frequency division multiplexing (OFDM) technique is applied, which employs $M$ subcarriers. Denoting $B$ and $f_c$ as the bandwidth and central frequency, respectively. The frequency on the $m$-th subcarrier is expressed as $f_m = f_c + \frac{B}{M} (m-1- \frac{M-1 }{2})$, $m = 1, \cdots M$. Due to the transmission characteristics of THz signals, line-of-sight (LoS) path is typically dominant. Therefore, we only consider the LoS path. In addition,
we assume the BS has full knowledge of the channel state information (CSI) for all links.
Consequently, the channel vector from $s$-th 6DMA of BS to the $k$-th user on the $m$-th subcarrier is denoted as
\begin{equation}
		\mathbf{h}_{s, m, k}=\eta_{m, k} \sqrt{g_k\left(\mathbf{u}_s\right)} \mathbf{a}\left(\mathbf{p}_s, \mathbf{u}_s, \theta_k, \varphi_k, f_m\right),
\end{equation}
where $\eta_{m, k}$ and $g_k\left(\mathbf{u}_s\right)$ represent the complex-valued path gain and effective antenna gain, respectively.   $\mathbf{a}\left(\mathbf{p}_s, \mathbf{u}_s, \theta_k, \varphi_k, f_m\right)$ denotes the array response vector of the $s$-th 6DMA surface. Specifically, the path gain is calculated as~\cite{refII_3}
\begin{eqnarray}
	\eta_{m, k}=\frac{\mathrm{c}}{4 \pi f_m d_{k}} e^{-\frac{1}{2} \kappa_{a b s}\left(f_m\right) d_{k}},
\end{eqnarray}
where $\kappa_{a b s}\left(f_m\right)$ indicates the molecular absorption factor and $d_{k}$
is the distance between the $k$-th user and BS. Let $\varphi_k \in [-\pi, \pi]$ and $\theta_k \in [0, \pi]$ represent the azimuth and elevation angles of a signal arriving at the $k$-th user, respectively. The array response vector is represent as
\begin{equation}\label{equ5}
\begin{aligned}
	&\mathbf{a}\left(\mathbf{p}_s, \mathbf{u}_s, \theta_k, \varphi_k, f_m\right)\\
	&=\left[e^{-j \frac{2 \pi f_m}{c} \mathbf{v}_k^T \mathbf{r}_{s, 1}\left(\mathbf{p}_s, \mathbf{u}_s\right)}, \ldots, e^{-j \frac{2 \pi f_m}{c} \mathbf{v}_k^T \mathbf{r}_{s, N}\left(\mathbf{p}_s, \mathbf{u}_s\right)}\right]^{T},
\end{aligned}
\end{equation}
where $c$ denotes the speed of light. $\mathbf{v}_k$ represents the pointing vector corresponding to the direction $(\theta_k, \varphi_k)$, which is
\begin{equation}
\mathbf{v}_k=[\sin(\theta_k)\cos(\varphi_k), \sin(\theta_k)\sin(\varphi_k), \cos(\varphi_k)]^T.
\end{equation}
Then, the effective antenna gain for the $s$-th 6DMA surface  is defined as
\begin{equation}
	g_k\left(\mathbf{u}_s\right)=G(\bar\theta_{s,k},\bar\varphi_{s,k})
\end{equation}
where 
\begin{equation}
	\begin{aligned}
		&\begin{aligned}
			& \bar{\theta}_{s,k}=\pi / 2-\arccos \left(\bar{z}_{s,k}\right), \\
			& \bar{\phi}_{s,k}=\arccos \left(\frac{\bar{x}_{s,k}}{\sqrt{\bar{x}_{s,k}^2+\bar{y}_{s,k}^2}}\right) \times \eta\left(\bar{y}_{s,k}\right),
		\end{aligned}\\
	\end{aligned}
\end{equation}
with $\eta\left(\bar{y}_{s,k}\right)=\left\{\begin{array}{c}
	1, \bar{y}_{s,k} \geq 0 \\
	-1, \bar{y}_{s,k}<0
\end{array}\right.$,  $\left[\bar{x}_{s,k}, \bar{y}_{s,k}, \bar{z}_{s,k}\right]^T=\mathbf{R}\left(\mathbf{u}_s\right)^{-1} \mathbf{v}_k$, and the function $G(\cdot)$ is determined by the radiation pattern of the deployed antenna~\cite{refI_19}.

\subsection{Beam Squint Effects of 6DMA}
For the above BS antenna architecture, directional beams are formed through a combination of analog and digital beamforming. However, conventional PSs are limited to generating frequency-independent phase shifts, whereas constructive interference across a wideband spectrum necessitates frequency-dependent phase shifts.
This inherent mismatch between the frequency-flat response of PSs and the frequency-selective requirement for constructive interference gives rise to beam squint effects. Consequently, beam energy becomes dispersed across subcarriers, and each subcarrier is directed towards a different spatial direction, reducing the transmission efficiency.

To analyze  the beam squint effects, we consider the array gain endowed by the 6DMA surface. For ease of expression, we consider single-user and single 6DMA surface scenario, and omit the subscript $k$ and $s$. If the frequency-independent PS-based analog beamforming
 $\mathbf{a}\left(\mathbf{p}, \mathbf{u}, \theta_0, \varphi_0, f_c\right)$ is employed~\cite{refII_4}, the normalized array gain on the $m$-th subcarrier at any direction $(\theta, \varphi)$ is calculated as
\begin{equation}
	\begin{aligned}
		\mathrm{g}\left(\mathbf{p}, \mathbf{u}, \theta, \varphi, f_m\right)&=\frac{1}{N}\left|\mathbf{a}^{H}\left(\mathbf{p}, \mathbf{u}, \theta_0, \varphi_0, f_c\right) \mathbf{a}\left(\mathbf{p}, \mathbf{u}, \theta, \varphi, f_m\right)\right| \\
		&=\frac{1}{N}\left|\sum_{n=1}^N e^{\frac{2 \pi}{c}\left(f_c \mathbf{v}_0^T-f_m \mathbf{v}^T\right) \mathbf{r}_{ n}\left(\mathbf{p}, \mathbf{u}\right)}\right|.
	\end{aligned}
\end{equation}
It can be observed that the array gain attains its maximum when $f_m = f_c$ and $\mathbf{v}_0 = \mathbf{v}$, which corresponds to a perfect alignment with the target location on the central subcarrier frequency. In contrast, for other subcarrier frequencies, such alignment cannot be preserved, resulting in an undesired beam deviation. Moreover, both the position and rotation $\mathbf{r}_{ n}\left(\mathbf{p}, \mathbf{u}\right)$ of the 6DMA are found to have an impact on the normalized array~gain.

\begin{figure}[t]
	\centering
	\includegraphics[height=2in,width=2.5in]{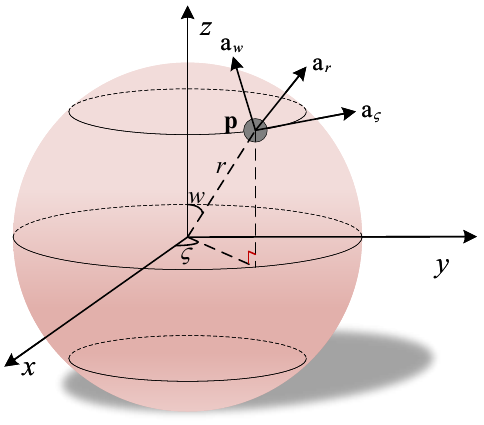}
	\caption{Illustration of the positions generated on a spherical surface.}
	\label{fig_2}
\end{figure}
Then, we further analyze the influence of the 6DMA position and rotation on the array gain. As illustrated in Fig. \ref{fig_2}, within region $\mathcal{D}$ of the 6DMA-BS area, a candidate position $\mathbf{p}$ is randomly selected from the spherical surface with the maximum radius~\cite{refII_5}. For the candidate position $\mathbf{p}$, its Cartesian coordinates are converted into spherical coordinates $(r, \omega, \zeta)$, where $r$, $\omega$, and $\zeta$ denote the radius, polar angle, and azimuth angle, respectively, relative to the BS reference position. Given $\mathbf{p}$, a unique rotation $\mathbf{u}$ is specified~\cite{refII_6}. In particular, the local $x'$-axis is oriented along the radial basis vector $\mathbf{c}_{r} = [\sin\omega \cos\zeta, \sin\omega \sin\zeta, \cos\omega]^T$, the $y'$-axis is oriented along the azimuthal basis vector $\mathbf{c}_{\zeta} = [-\sin\zeta, \cos\zeta, 0]^T$, and the $z'$-axis is aligned with the polar basis vector $\mathbf{c}_{\omega} = [\cos\omega \cos\zeta, \cos\omega \sin\zeta, -\sin\omega]^T$. Consequently, the rotation matrix at the position $\mathbf{p}$ is given by $\mathbf{R}\left(\mathbf{u}\right)=[\mathbf{c}_{r}, \mathbf{c}_{\zeta}, \mathbf{c}_{\omega}]$. Based on $\mathbf{R}(\mathbf{u})$ and $\mathbf{p}$, the position $\mathbf{r}_{n}(\mathbf{p}, \mathbf{u})$ of the $n$-th element of the 6DMA surface in the global CCS can be determined.

\begin{figure}[t]
	\centering
	\subfloat[]{\includegraphics[height=2.2in,width=3.2in]{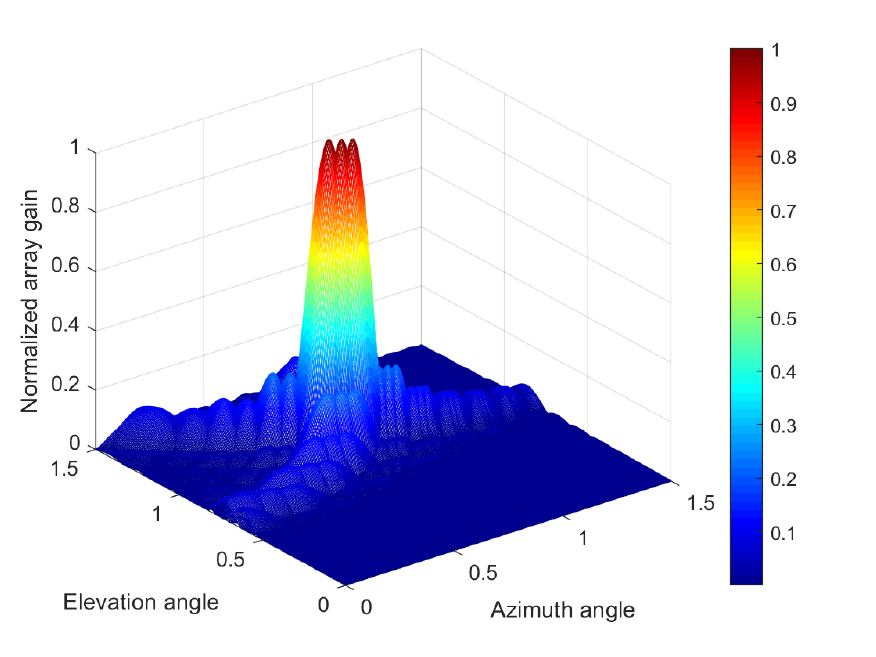}%
		\label{fig_3a}}
	\hfil
	\subfloat[]{\includegraphics[height=2.2in,width=3.2in]{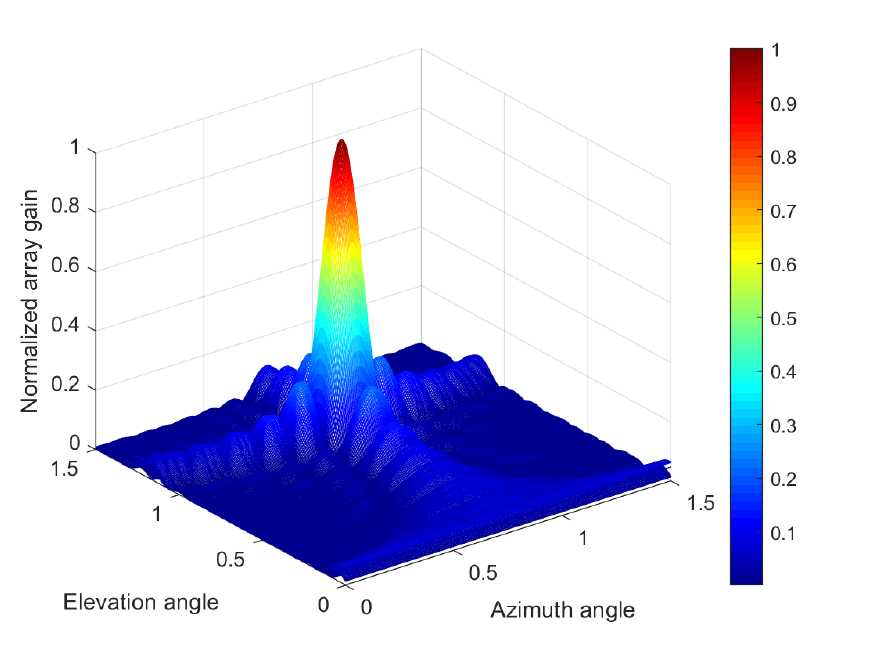}%
		\label{fig_3b}}
	\caption{Beam pattern. (a) Normal vector deviates from the user.  (b) Normal vector points towards the user.}
	\label{fig_3}
\end{figure}

\begin{figure}[htbp]
	\centering
	\includegraphics[height=2.2in,width=3in]{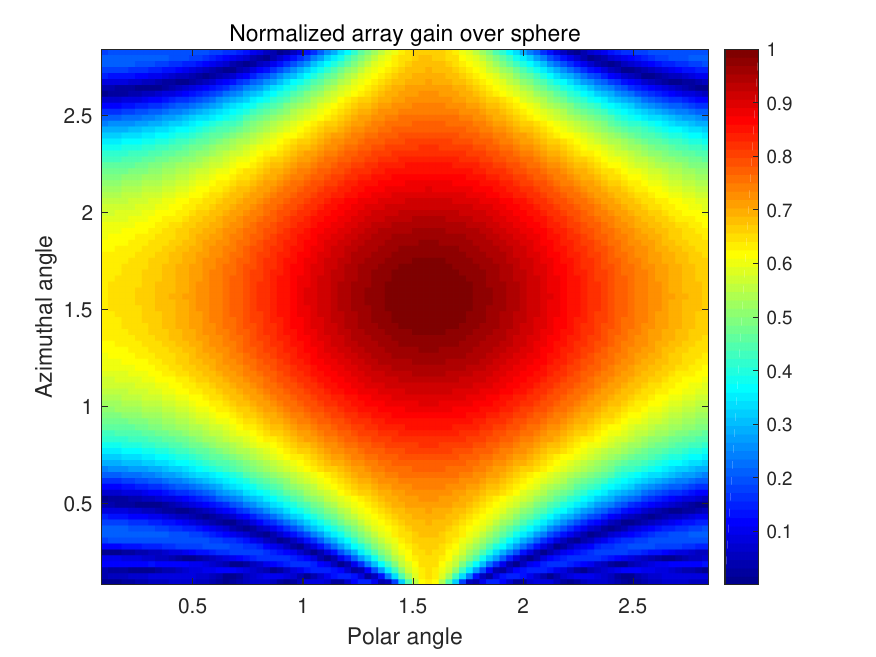}
	\caption{The normalized array gain across different positions on the spherical surface.}
	\label{fig_4}
\end{figure}

To illustrate the beam squint effects, Fig. \ref{fig_3} presents the normalized array gain achieved through frequency-independent beamforming vectors at the center frequency $f_c$ and two edge subcarriers $f_1$ and $f_M$, where $f_c = 300$ GHz and $B=20$ GHz. The elevation and azimuth angles are adopted as $\theta=\pi/3$ and $\varphi=\pi/4$, respectively. In Fig. \ref{fig_3} (a), the 6DMA surface is placed at the spherical coordinate $\mathbf{p}=(0.5, \pi/4, \pi/2)$, and its normal vector deviates from the user's direction. In Fig. \ref{fig_3} (b), the 6DMA surface is located at $\mathbf{p}=(0.5, \pi/3, \pi/4)$,  and its normal vector points towards the user direction. The results indicate that the 6DMA locations significantly influence the severity of beam squint. Fig. \ref{fig_4} further evaluates the normalized array gain at the edge frequency $f_M$ across different positions on the spherical surface, with the user located along the $y$-axis. It can be found that the array gain reaches its maximization when the 6DMA surface normal vector is closely aligned with the user direction. 

Based on the above analysis, unlike conventional fixed-position antenna  architecture that require costly TTD for overcoming beam squint effects, 6DMA offers a cost-efficient scheme by reconfiguring the array geometry to realize the beam alignment across subcarriers.

\section{Problem Formulation and Solution}
In this section, we first formulate a sum-rate maximization problem under 6DMA-enabled wideband THz communication systems. Then, we propose an AO algorithm to efficiently solve this problem.

\subsection{Problem Formulation}
The received signal of the $k$-th user on the $m$-th subcarrier can be expressed as
\begin{align}
	y_{m, k}=\mathbf{h}_{m, k} \mathbf{A} \mathbf{d}_{m, k} s_{m, k}+\sum_{j=1, j \neq k}^K \mathbf{h}_{m, k} \mathbf{A} \mathbf{d}_{m, j} s_{m, j}+\mathrm{n}_{m, k},
\end{align}
where $\mathbf{h}_{m, k}=[\mathbf{h}_{1, m, k}, \mathbf{h}_{2, m, k}, \cdots, \mathbf{h}_{S, m, k}]$, $\mathbf{A}=\mathrm{blkdiag}(\mathbf{b}_1, \mathbf{b}_2, \cdots, \mathbf{b}_{N_{\rm RF}},)\in \mathbb{C}^{N_t \times N_{\rm{RF}}}$ is analog beamforming matrix with $\mathbf{b}_l=[e^{j\phi_{(l-1)N+1}}, \cdots, e^{j\phi_{lN}}]^T \in \mathbb{C}^{N \times 1}$,  
$\mathbf{d}_{m, k}\in \mathbb{C}^{N_{\rm{RF}} \times 1}$ is the digital beamforming vector, $s_{m, k}$ is the transmit symbol for the $k$-th user on the $m$-th subcarrier, and $n_{m,k} \sim \mathcal{C} \mathcal{N}\left(0, \sigma_{m,k}^{2}\right)$ represents the additive white Gaussian noise (AWGN). Let $\mathbf{s}_m=[\mathbf{s}_{m, 1}, \cdots,\mathbf{s}_{m, k}] \in \mathbb{C}^{K\times 1}$ denote the transmit symbol vector on the $m$-th subcarrier, which satisfies  $\mathbb{E}\left[\mathbf{s}_m\mathbf{s}_m^H\right]=\frac{P}{K}\mathbf{I}_K$ by assuming equal power allocation among the users. Note that $P$ indicates the average power. Thus, we have
the total transmit power constraint as $\left\|\mathbf{A}\mathbf{D}_{m}\right\|_F^{2}=K$ with $\mathbf{D}_m=[\mathbf{d}_{m, 1}, \cdots,\mathbf{d}_{m, k}] \in \mathbb{C}^{N_{\rm{RF}} \times K}$. 

Then, the $k$-th user's SINR, $ \forall k \in \mathcal{K}$, on the $m$-th subcarrier is computed by
\begin{eqnarray}
	\gamma_{m,k}=\frac{\frac{P}{K}\left|\mathbf{h}_{m, k} \mathbf{A} \mathbf{d}_{m, k}\right|^{2}}{\frac{P}{K} \sum_{j=1, j \neq k}^{K}\left|\mathbf{h}_{m, k} \mathbf{A} \mathbf{d}_{m, j}\right|^{2}+\sigma_{m, k}^{2}},
\end{eqnarray}
and the corresponding sum-rate is
\begin{eqnarray}
	R_{\rm{sum}}=\sum_{k=1}^{K} \sum_{m=1}^{M} \log _{2}\left(1+\gamma_{m, k}\right).
\end{eqnarray}
Finally, the sum-rate maximization problem can be expressed~as
\begin{subequations}\label{Opt_A}
	\begin{align}
		\mathrm{P0}:&\max _{\mathbf{A}, \mathbf{D}_{m}, \mathbf{p}, \mathbf{u}} R_{\rm{sum}}\label{Opt_A0}\\
		{\rm{s.t.}}\;\;&\mathbf{p}_i \in \mathcal{D}
		 \label{Opt_A1},\\
		&\left|\mathbf{b}_{l, n}\right|=1, l=1, \ldots N_{\rm{RF}}, n=1,\ldots, N\label{Opt_A2},\\
		&\left\|\mathbf{A}\mathbf{D}_{m}\right\|_F^{2}=K \label{Opt_A3},\\
		&\left\|\mathbf{p}_i-\mathbf{p}_j\right\|_2 \geq d_{\min }, \forall i, j \in S, j \neq i \label{Opt_A4},\\
		&\mathbf{n}\left(\mathbf{u}_i\right)^T\left(\mathbf{p}_j-\mathbf{p}_i\right) \leq 0, \forall i, j \in S, j \neq i \label{Opt_A5}, \\
		& \mathbf{n}\left(\mathbf{u}_i\right)^T \mathbf{p}_i \geq 0, \forall i \in S \label{Opt_A6},
	\end{align}
\end{subequations}
where $\mathbf{n}\left(\mathbf{u}_i\right) = \mathbf{R}\left(\mathbf{u}_i\right) \bar{\mathbf{n}}$ and $\bar{\mathbf{n}}$ denotes the normal vector of the $i$-th 6DMA surface in the local CCS. Constraint \eqref{Opt_A1} ensures that all 6DMA surfaces remain within the feasible 3D deployment region $\mathcal{D}$. Constraint \eqref{Opt_A2} imposes the unit-modulus requirement induced by PSs. \eqref{Opt_A3} is the transmit power constraint. Constraint \eqref{Opt_A4} guarantees a minimum inter-surface distance $d_{\min}$ to avoid physical overlap, and \eqref{Opt_A5} prevents undesired mutual reflections between any two surfaces. Furthermore, \eqref{Opt_A6} restricts the surface rotation to avoid orientations towards the CPU. Due to the presence of coupled variables, the non-concave objective function in \eqref{Opt_A0}, and the non-convex constraints \eqref{Opt_A2}–\eqref{Opt_A6}, solving the formulated problem becomes challenging. In the next subsection, we develop an efficient algorithmic framework to address it.

\subsection{Problem Solution}
 In this section, we design an AO scheme to iteratively address $\mathbf{A}$, $\mathbf{D}_{m}$, $\mathbf{p}$, and $\mathbf{u}$. Specifically, given the positions $\mathbf{p}$ and rotations $\mathbf{u}$, the analog beamforming matrix $\mathbf{A}$ and digital beamforming matrices $\mathbf{D}_{m}$ are first obtained using a SDR based alternating minimization algorithm. Subsequently, with $\mathbf{A}$ and $\mathbf{D}_{m}$ fixed, the positions and rotations of all 6DMA surfaces are updated via the feasible gradient descent method. The final solution is obtained by iteratively repeating the above steps until convergence.

\subsubsection{Hybrid Beamforming Design}
Given positions $\mathbf{p}$ and rotations $\mathbf{u}$ of all 6DMA surfaces, the sum-rate maximization problem P0 in \eqref{Opt_A} can be reformulated as the  following subproblem P1 for the analog and digital beamforming design
\begin{subequations}\label{Opt_B}
	\begin{align}
		\mathrm{P1}:&\max _{\mathbf{A}, \mathbf{D}_{m}} R_{\rm{sum}}\label{Opt_B0}\\
		{\rm{s.t.}}\;\;
		&\eqref{Opt_A2}, \eqref{Opt_A3}\label{Opt_B1}.
	\end{align}
\end{subequations}

However, due to the extremely complex mathematical form of $R_{\rm{sum}}$, the direct optimization of analog and digital beamforming is challenging.  It has been theoretically proved \cite{refII_4} that the problem of maximizing the sum-rate can be approximated as a matrix factorization problem. Thus, P1 can be further reformulated as
\begin{subequations}\label{Opt_D}
	\begin{align}
		\mathrm{P1-A}:&\min _{\mathbf{A}, \mathbf{D}_{m}} \sum_{m=1}^{M}\left\|\mathbf{F}_m^{\rm opt}-\mathbf{A}\mathbf{D}_{m}\right\|_F^{2}\label{Opt_C0}\\
		{\rm{s.t.}}\;\;
		&\eqref{Opt_A2}, \eqref{Opt_A3}\label{Opt_C1},
	\end{align}
\end{subequations}
where $\mathbf{F}_m^{\rm opt}$ denotes the unconstrained optimal beamforming matrix at the $m$-th subcarrier. 
Accordingly, the beamforming design problem is reformulated as the joint optimization of the digital beamforming matrix $\mathbf{D}_{m}$ and the analog beamforming matrix $\mathbf{A}$, such that their product closely approximates $\mathbf{F}_m^{\rm opt}$ across all subcarriers.

On account of the special form  of the constraint on the matrix  $\mathbf{A}$, each nonzero element of $\mathbf{A}$ scales the corresponding row of $\mathbf{D}_{m}$ in the product $\mathbf{A}\mathbf{D}_{m}$. Consequently, the transmit power constraint in \eqref{Opt_A3} can be equivalently expressed as
\begin{align}\label{relation}
 \left\|\mathbf{A}\mathbf{D}_{m}\right\|_F^{2}=\frac{N_t}{N_{\rm NF}}\left\|\mathbf{D}_{m}\right\|_F^{2}.
\end{align}
Based on \eqref{relation}, the analog beamforming design can be formulated as
\begin{subequations}\label{Opt_C}
	\begin{align}
		\mathrm{P1-B}:&\min _{\mathbf{A}} \sum_{m=1}^{M}\left\|\mathbf{F}_m^{\rm opt}-\mathbf{A}\mathbf{D}_{m}\right\|_F^{2}\label{Opt_D0}\\
		{\rm{s.t.}}\;\;
		&\eqref{Opt_A2}\label{Opt_D1}.
	\end{align}
\end{subequations}
Moreover, due to the same property of $\mathbf{A}$, P1-B is further formulated as
\begin{align}
\min _{\{\phi_\iota\}_{\iota=1}^{N_t}} \sum_{m=1}^{M}\left\|\left(\mathbf{F}_m^{\rm opt}\right)_{\iota,:}-e^{j\phi_\iota}\left(\mathbf{D}_{m}\right)_{q,:}\right\|_2^{2},
\end{align}
where $q=\lceil \iota \frac{N_t}{N_{\rm NF}} \rceil$. This problem can be interpreted as a vector approximation task via phase rotation, and there holds a closed-form expression for nonzero elements in $\mathbf{A}$, which is expressed~as
\begin{align}\label{eqn_16}
	\mathrm{arg}\{\left(\mathbf{A}\right)_{\iota,q}\}&=\mathrm{arg}\{\sum_{m=1}^{M}\left(\mathbf{F}_m^{\rm opt}\right)_{\iota,:}\left(\mathbf{D}_m\right)_{q,:}^H\},\notag\\
	&1\leq \iota \leq N_t, q=\lceil \iota \frac{N_t}{N_{\rm RF}} \rceil.
\end{align}
Once the solution for analog beamforming is obtained, we then design the digital beamforming, and P1-A can be further simplified as
\begin{subequations}\label{Opt_E}
	\begin{align}
		\mathrm{P1-C}:&\min _{\mathbf{D}_{m}} \sum_{m=1}^{M}\left\|\mathbf{F}_m^{\rm opt}-\mathbf{A}\mathbf{D}_{m}\right\|_F^{2}\label{Opt_E0}\\
		{\rm{s.t.}}\;\;
		&\left\|\mathbf{D}_{m}\right\|_F^{2}=\frac{K N_{\rm NF}}{N_t}   \label{Opt_E1}.
	\end{align}
\end{subequations}
P1-C is a non-convex quadratic constraint quadratic programming (QCQP) problem, which can be further stated  as a homogeneous QCQP problem:
\begin{subequations}\label{Opt_F}
	\begin{align}
	\mathrm{P1-D}:	&\underset{\mathbf{X} \in \mathbb{H}^u}{\operatorname{min}}  \operatorname{Tr}(\mathbf{C X}) \label{Opt_F0}\\
		{\rm{s.t.}}\;\;& 
			\operatorname{Tr}\left(\mathbf{B}_1 \mathbf{X}\right)=\frac{K  N_{\mathrm{RF}} }{N_t}, \label{Opt_F1}\\&
			\operatorname{Tr}\left(\mathbf{B}_2 \mathbf{X}\right)=1, \label{Opt_F2}\\&
			\mathbf{X} \succeq 0, \label{Opt_F3}\\&
			\operatorname{rank}(\mathbf{X})=1, \label{Opt_F4}
	\end{align}
\end{subequations}
where $\mathbb{H}^u$ is the set of $u=K N_{\mathrm{RF}}+1$ dimension complex Hermitian matrices. In addition, 
\begin{equation}
	\begin{aligned}
		& \mathbf{X}=\mathbf{x x}^H, \mathbf{x}=\left[\begin{array}{c}
			\mathbf{b} \\
			t
		\end{array}\right],  \mathbf{C}=\left[\begin{array}{cc}
			\mathbf{E}^H \mathbf{E} & -\mathbf{E}^H \mathbf{f}_m \\
			-\mathbf{f}_m^H \mathbf{E} & \mathbf{f}_m^H \mathbf{f}_m
		\end{array}\right], \\
	\end{aligned}
\end{equation}

\begin{equation}
	\begin{aligned}
		& \mathbf{B}_1=\left[\begin{array}{cc}
			\mathbf{I}_{U-1} & \mathbf{0} \\
			\mathbf{0} & 0
		\end{array}\right], \mathbf{B}_2=\left[\begin{array}{cc}
			\mathbf{0}_{U-1} & \mathbf{0} \\
			\mathbf{0} & 1
		\end{array}\right],
	\end{aligned}
\end{equation}
where $\mathbf{b}=\rm{vec}(\mathbf{D}_{m})$,  $\mathbf{E}=\mathbf{I}_{K} \otimes \mathbf{F}$, $\mathbf{f}_m=\text{vec}(\mathbf{F}_m^{\rm opt})$, and $t$ is an auxiliary variable.

The major challenge in solving problem P1-D lies in the rank-one constraint, which is non-convex relative to $\mathbf{X}$. To address this, we first relax the constraint and obtain a simplified version of P1-D, forming a standard SDR problem.
\begin{subequations}\label{Opt_G}
	\begin{align}
		\mathrm{P1-E}:	&\underset{\mathbf{X} \in \mathbb{H}^u}{\operatorname{min}}  \operatorname{Tr}(\mathbf{C X}) \label{Opt_G0}\\
		{\rm{s.t.}}\;\;& 
		\eqref{Opt_F1}, \eqref{Opt_F2}, \eqref{Opt_F3} . 
	\end{align}
\end{subequations}
It is well established that for a homogeneous QCQP with complex values, the SDR is compact  when the number of constraints is fewer than three \cite{refII_7}. Hence, by dropping the rank-one constraint, problem P1-E transforms into a SDR problem that can be efficiently solved using standard convex optimization algorithms \cite{refII_8}. This yields the globally optimal solution for designing the digital beamforming \eqref{Opt_E}. The overall SDR-based hybrid beamforming optimization framework is summarized in \textbf{Algorithm 1}.

\begin{algorithm}[t]
	\caption{SDR-based Algorithm for Solving $\mathrm{P1}$}
	{\bf Input:}
	Optimal beamforming matrix $\mathbf{F}_m^{\rm opt}$.\\
	{\bf Initialization:}
	Analog beamforming matrix $\mathbf{A}^{(0)}$.\\
	{\bf while} $0 \leq i \textless I_1$ {\bf do}\\
	Derive the digital beamforming  $\mathbf{D}_{m}$ by solving P1-E;\\
	Compute the analog beamforming  $\mathbf{A}$ by \eqref{eqn_16}.\\
	{\bf end while}
	
	{\bf Output:}
	Analog beamforming matrix $\mathbf{A}$, digital  beamforming matrix $\mathbf{D}_{m}$.
\end{algorithm}

\subsubsection{Positions and Rotations Optimization}
After obtaining $\mathbf{A}$ and $\mathbf{D}_{m}$, we optimize positions $\mathbf{p}$ and rotations $\mathbf{u}$ of all 6DMA surfaces. The sum-rate maximization problem  P0 in \eqref{Opt_A}  is reformulated as follows.
\begin{subequations}\label{Opt_H}
	\begin{align}
		\mathrm{P2}:&\max _{\mathbf{p}, \mathbf{u}} R_{\rm{sum}}\label{Opt_H0}\\
		{\rm{s.t.}}\;\;&\eqref{Opt_A1}, \eqref{Opt_A4},\eqref{Opt_A5},\eqref{Opt_A6}.
	\end{align}
\end{subequations}
During the optimization process of $\mathbf{p}$ and $\mathbf{u}$, the alternating iterative method is applied. In each iteration, we first solve $\mathbf{p}_s$ with given $\mathbf{p}_j$, $j \in S, j \neq s$ and  $\mathbf{u}_j$, $j \in S$. Thus, the problem for designing $\mathbf{p}_s$ can be written as
\begin{subequations}\label{Opt_I}
	\begin{align}
		\mathrm{P2-A}:&\max _{\mathbf{p}_s} R_{\rm{sum}}\label{Opt_I0}\\
		{\rm{s.t.}}\;\;&\mathbf{p}_s \in \mathcal{D}
		\label{Opt_I1},\\
		&\left\|\mathbf{p}_s-\mathbf{p}_j\right\|_2 \geq d_{\min }, \forall s, j \in S, j \neq s \label{Opt_I2},\\
		&\mathbf{n}\left(\mathbf{u}_s\right)^T\left(\mathbf{p}_j-\mathbf{p}_s\right) \leq 0, \forall s, j \in S, j \neq s \label{Opt_I3}, \\
	&\mathbf{n}\left(\mathbf{u}_j\right)^T\left(\mathbf{p}_s-\mathbf{p}_j\right) \leq 0, \forall s, j \in S, j \neq s \label{Opt_I4}, \\
		& \mathbf{n}\left(\mathbf{u}_s\right)^T \mathbf{p}_s \geq 0, \forall s \in S \label{Opt_I5}.
	\end{align}
\end{subequations}
Due to the non-convex constraint \eqref{Opt_I2}, we first convert it into a convex form. According to \cite{refI_19}, \eqref{Opt_I2} can be approximated as the following linear inequality,
\begin{equation}\label{linear inequality}
	\left(\mathbf{p}_j-\mathbf{p}_s^{(\kappa-1)}\right)^T\left(\mathbf{p}_s-\mathbf{p}_{\text {boun}, j}\right) \leq 0, \forall s, j \in S, j \neq s,
\end{equation}
where $\mathbf{p}_s^{(\kappa-1)}$ is the value of $\mathbf{p}_s$ in the $(\kappa-1)$-th iteration. In addition,
\begin{equation}\label{boundary}
\mathbf{p}_{\text {boun}, j}= \mathbf{p}_j- \frac{d_{\min }}{\left\|\mathbf{p}_j-\mathbf{p}_s^{(\kappa-1)}\right\|_2}(\mathbf{p}_j-\mathbf{p}_s^{(\kappa-1)}), \forall s, j \in S, j \neq s.
\end{equation}
Therefore, P2-A is further transformed as
\begin{subequations}\label{Opt_J}
 	\begin{align}
 		\mathrm{P2-B}:&\max _{\mathbf{p}_s} R_{\rm{sum}}\label{Opt_J0}\\
 		{\rm{s.t.}}\;\;&\eqref{Opt_I1}, \eqref{Opt_I3}, \eqref{Opt_I4},\eqref{Opt_I5},\eqref{linear inequality}.
 	\end{align}
\end{subequations}
The feasible region with respect to $\mathbf{p}_s$ becomes a convex set. Then, we apply the feasible gradient descent method to solve the above problem. Specifically, the feasible gradient descent method starts with a feasible vector $\mathbf{p}_s^{(\kappa-1)}$  and generates another feasible vector $\mathbf{p}_s^{(\kappa)}$  as 
\begin{equation}\label{feasible_p}
\mathbf{p}_s^{(\kappa)}=\mathbf{p}_s^{(\kappa-1)}+\tau^{(\kappa-1)}(\bar{\mathbf{p}}_s^{(\kappa-1)}-\mathbf{p}_s^{(\kappa-1)}),
\end{equation}
where $\tau^{(\kappa-1)} \in(0,1]$ is the adaptive step size computed by the Armijo rule \cite{refII_9}, $\bar{\mathbf{p}}_s^{(\kappa-1)}$ is a feasible vector , and $\bar{\mathbf{p}}_s^{(\kappa-1)}-\mathbf{p}_s^{(\kappa-1)}$ is a feasible direction. In particular, $\bar{\mathbf{p}}_s^{(\kappa-1)}$ is selected as the
solution to the optimization problem below.
\begin{subequations}\label{Opt_K}
	\begin{align}
		\mathrm{P2-C}:&\min _{\mathbf{p}_s} - \nabla_{\mathbf{p}_s} f\left(\mathbf{p}_s^{(\kappa-1)}, \mathbf{u}_s\right)^T (\mathbf{p}_s-\mathbf{p}_s^{(\kappa-1)})  \label{Opt_K0}\\
		{\rm{s.t.}}\;\;&\eqref{Opt_I1}, \eqref{Opt_I3}, \eqref{Opt_I4},\eqref{Opt_I5},\eqref{linear inequality},
	\end{align}
\end{subequations}
where the gradient of function $\nabla_{\mathbf{p}_s} f\left(\mathbf{p}_s^{(\kappa-1)}, \mathbf{u}_s\right)$ can be derived by 
\begin{equation}\label{gradient}
	\begin{aligned}
		& {\left[\nabla_{\mathbf{p}_s} f\left(\mathbf{p}_s^{(\kappa-1)}, \mathbf{u}_s\right)\right]_j} \\
		& =\lim _{\epsilon \rightarrow 0} \frac{f\left(\mathbf{p}_s^{(\kappa-1)}+\epsilon \mathbf{e}^j, \mathbf{u}_s\right)-f\left(\mathbf{p}_s^{(\kappa-1)}, \mathbf{u}_s\right)}{\epsilon}, \\
		& \quad 1 \leq j \leq 3,
	\end{aligned}
\end{equation}
where $\mathbf{e}^j \in \mathbb{R}^3$ denotes a three-dimensional vector whose $j$-th element equals 1, while all other elements are 0. It is worth noting that problem P2-C is a linear optimization problem that can be addressed effectively by adopting linprog \cite{refII_9}. The overall procedure for tackling  problem P2-A is presented in \textbf{Algorithm 2}.
\begin{algorithm}[t]
	\caption{Feasible Gradient Descent Algorithm for Solving P2-A}
	{\bf Input:} Step size $\tau_{\mathrm{ini}}$, $\mu = 10^{-2}$, $\delta = 0.5$, $\{\mathbf{p}_j\}_{j \in \mathcal{S}/s}$, $\{\mathbf{u}_{j}\}_{j \in \mathcal{S}}$, and  $\mathbf{u}_{s}^{(0)}$.\\ 
	{\bf Initialization:}
	$\kappa \gets 0$.\\
	{\bf while} $0 \leq i \textless I_2$ {\bf do}\\
		Compute $\nabla_{\mathbf{p}_s} f(\mathbf{p}_s^{(\kappa-1)}, \mathbf{u}_s)$ based on \eqref{gradient} and
		set $\tau \gets \tau_{\mathrm{ini}}$;\\
		Obtain $\bar{\mathbf{p}}_s^{(\kappa-1)}$ by solving P2-C;\\
		Compute $\mathbf{p}_s^{(\kappa)} \gets \mathbf{p}_s^{(\kappa-1)} + \tau \left( \bar{\mathbf{p}}_s^{(\kappa-1)} - \mathbf{p}_s^{(\kappa-1)} \right)$;\\
		 \hspace*{1em}{\bf while} $f(\mathbf{p}_s^{(\kappa)}, \mathbf{u}_s) - f(\mathbf{p}_s^{(\kappa-1)}, \mathbf{u}_s) < \hspace*{3em} \mu \tau \nabla_{\mathbf{p}_s} f(\mathbf{p}_s^{(\kappa-1)}, \mathbf{u}_s)^{T} (\bar{\mathbf{p}}_s^{(\kappa-1)} - \mathbf{p}_s^{(\kappa-1)})$ {\bf do}\\
		\hspace*{1em} $\tau \gets \delta \tau$;\\
		\hspace*{1em}	Compute $\mathbf{p}_s^{(\kappa)} \gets \mathbf{p}_s^{(\kappa-1)} + \tau \left( \bar{\mathbf{p}}_s^{(\kappa-1)} - \mathbf{p}_s^{(\kappa-1)} \right)$;\\
		\hspace*{1em} {\bf end while}\\
		Update $\kappa \gets \kappa + 1$;\\
		{\bf end while}\\
	{\bf Return} $\mathbf{p}_s$.
\end{algorithm}

Next, we design $\mathbf{u}_s$ with given $\mathbf{u}_j$, $j \in S, j \neq s$ and  $\mathbf{p}_j$, $j \in S$. The problem for optimizing $\mathbf{u}_s$ can be further formulated as follows.
\begin{subequations}\label{Opt_L}
	\begin{align}
		\mathrm{P3-A}:&\max _{\mathbf{u}_s} R_{\rm{sum}}\label{Opt_L0}\\
		{\rm{s.t.}}\;\;
		&\mathbf{n}\left(\mathbf{u}_s\right)^T\left(\mathbf{p}_j-\mathbf{p}_s\right) \leq 0, \forall s, j \in S, j \neq s \label{Opt_L1}, \\
		& \mathbf{n}\left(\mathbf{u}_s\right)^T \mathbf{p}_s \geq 0, \forall s \in S \label{Opt_L2}.
	\end{align}
\end{subequations}
Similarly, due to the non-convex constraints \eqref{Opt_L1} and \eqref{Opt_L2}, we first transform these constraints into convex form. We denote $\mathbf{u}_s^{(\kappa-1)}=[\alpha_s^{(\kappa-1)}, \beta_s^{(\kappa-1)}, \gamma_s^{(\kappa-1)}]^T  $ as the rotation value after the $(\kappa-1)$-th iteration 
and $\Delta \mathbf{u}_s=\mathbf{u}_s-\mathbf{u}_s^{(\kappa-1)}=[\Delta \alpha_s, \Delta \beta_s, \Delta \gamma_s]^T$ as the corresponding increments in the $\kappa$-th iteration, where $\Delta \alpha_s=\alpha_s-\alpha_s^{(\kappa-1)}$, $\Delta \beta_s=\beta_s-\beta_s^{(\kappa-1)}$, and $\Delta \gamma_s=\gamma_s-\gamma_s^{(\kappa-1)}$. The update  of the rotation matrix at the current iteration $\mathbf{R}(\mathbf{u}_s)$ is calculated by
\begin{equation}\label{incremental}
\mathbf{R}(\mathbf{u}_s)=\mathbf{R}(\mathbf{u}_s^{(\kappa-1)})  \mathbf{R}(\Delta \mathbf{u}_s),
\end{equation}
where $\mathbf{R}(\mathbf{u}_s^{(\kappa-1)})$ is the rotation matrix at the $(\kappa-1)$-th iteration and $ \mathbf{R}(\Delta \mathbf{u}_s)$ is the incremental rotation  matrix. Since the increments $\Delta \mathbf{u}_s$ are very small in each iteration, we can take the following small-angle approximations: 1) $ \cos(x)\rightarrow 1 $ and 2) $ \sin(x) \rightarrow x $, when $x \rightarrow 0$. By following the  above approximations, the incremental rotation matrix $\mathbf{R}(\Delta \mathbf{u}_s)$  can be approximated as
\begin{equation}\label{rotation}
	\mathbf{R}\left(\Delta \mathbf{u}_s\right) \approx\left[\begin{array}{ccc}
		1 & \Delta \gamma_s & -\Delta \beta_s \\
		-\Delta \gamma_s & 1 & \Delta \alpha_s \\
		\Delta \beta_s & -\Delta \alpha_s & 1
	\end{array}\right].
\end{equation}
By substituting \eqref{incremental} and \eqref{rotation} into $\mathbf{n}\left(\mathbf{u}_s\right) = \mathbf{R}\left(\mathbf{u}_s\right) \bar{\mathbf{n}}$, 
the constraints \eqref{Opt_L1} and \eqref{Opt_L2} can be respectively linearized as
\begin{equation}\label{constraint1}
\bar{\mathbf{n}}^T \mathbf{R}\left(\Delta \mathbf{u}_s\right)^T \mathbf{R}(\mathbf{u}_s^{(\kappa-1)})^T \left(\mathbf{p}_j-\mathbf{p}_s\right) \leq 0, \forall s, j \in S, j \neq s, 
\end{equation}
\begin{equation}\label{constraint2}
\bar{\mathbf{n}}^T \mathbf{R}\left(\Delta \mathbf{u}_s\right)^T \mathbf{R}(\mathbf{u}_s^{(\kappa-1)})^T\left(\mathbf{u}_s\right)^T \mathbf{p}_s \geq 0, \forall s \in S.
\end{equation}
Thus, P3-A can be transformed as
\begin{subequations}\label{Opt_M}
	\begin{align}
		\mathrm{P3-B}:&\max _{\mathbf{p}_s} R_{\rm{sum}}\label{Opt_M0}\\
		{\rm{s.t.}}\;\;&\eqref{constraint1}, \eqref{constraint2}.
	\end{align}
\end{subequations}
The feasible region with respect to $\mathbf{u}_s$ becomes a convex set. Then, we still apply the feasible gradient descent method to solve this problem. Specifically, the feasible gradient descent method starts with a feasible vector $\mathbf{u}_s^{(\kappa-1)}$  and produces another feasible vector $\mathbf{u}_s^{(\kappa)}$  as 
\begin{equation}\label{feasible_u}
	\mathbf{u}_s^{(\kappa)}=\mathbf{u}_s^{(\kappa-1)}+\tau^{(\kappa-1)}(\bar{\mathbf{u}}_s^{(\kappa-1)}-\mathbf{u}_s^{(\kappa-1)}),
\end{equation}
where $\bar{\mathbf{u}}_s^{(\kappa-1)}$ is a feasible vector and $\bar{\mathbf{u}}_s^{(\kappa-1)}-\mathbf{u}_s^{(\kappa-1)}$ is a feasible direction. In particular, $\bar{\mathbf{u}}_s^{(\kappa-1)}$ can be chosen as the
solution to the following optimization problem.
\begin{subequations}\label{Opt_N}
	\begin{align}
		\mathrm{P3-C}:&\min _{\mathbf{u}_s} -\nabla_{\mathbf{u}_s} f\left(\mathbf{p}_s, \mathbf{u}_s^{(\kappa-1)}\right)^T (\mathbf{u}_s-\mathbf{u}_s^{(\kappa-1)})  \label{Opt_N0}\\
		{\rm{s.t.}}\;\;&\eqref{constraint1}, \eqref{constraint2},
	\end{align}
\end{subequations}
where the gradient of function $\nabla_{\mathbf{u}_s} f\left(\mathbf{p}_s, \mathbf{u}_s^{(\kappa-1)}\right)$ can be derived by 
\begin{equation}\label{gradient_u}
	\begin{aligned}
		& {\left[\nabla_{\mathbf{u}_s} f\left(\mathbf{p}_s, \mathbf{u}_s^{(\kappa-1)}\right)\right]_j} \\
		& =\lim _{\epsilon \rightarrow 0} \frac{f\left(\mathbf{p}_s, \mathbf{u}_s^{(\kappa-1)}+\epsilon \mathbf{e}^j\right)-f\left(\mathbf{p}_s, \mathbf{u}_s^{(\kappa-1)}\right)}{\epsilon}, \\
		& \quad 1 \leq j \leq 3,
	\end{aligned}
\end{equation}
where $\mathbf{e}^j \in \mathbb{R}^3$ denotes a three-dimensional vector whose $j$-th element equals 1, while all other elements are 0. It is worth noting that problem P3-C is a linear optimization problem that can be addressed effectively by adopting linprog. The overall steps for solving problem P3-A is presented in~\textbf{Algorithm 3}.

So far, the analog beamforming matrix $\mathbf{A}$, digital beamforming matrix $\mathbf{D}_{m}$, positions $\mathbf{p}$, and rotations $\mathbf{u}$ have all been obtained. For solving P0, given positions $\mathbf{p}$ and rotations $\mathbf{u}$, we first design the SDR-based alternating minimization algorithm to obtain the hybrid beamforming, including analog beamforming $\mathbf{A}$ and digital beamforming $\mathbf{D}_{m}$. After that,  the optimization is decoupled into two subproblems: 1) the positions optimization and 2) the rotations optimization, which are alternately solved based on designing the feasible gradient descent method. The above steps are repeated until convergence. The details of the proposed AO algorithm for solving problem P0 are summarized in \textbf{Algorithm~4}. 

\begin{algorithm}[t]
	\caption{Feasible Gradient Descent Algorithm for Solving P3-A}
	{\bf Input:}
	Step size $\tau_{\mathrm{ini}}$, $\mu = 10^{-2}$, $\delta = 0.5$, $\{\mathbf{u}_j\}_{j \in \mathcal{S}/s}$, $\{\mathbf{p}_{j}\}_{j \in \mathcal{S}}$, and  $\mathbf{p}_{s}^{(0)}$.\\ 
	{\bf Initialization:}
	$\kappa \gets 0$.\\
	{\bf while} $0 \leq i \textless I_3$ {\bf do}\\
	Compute $\nabla_{\mathbf{u}_s} f(\mathbf{p}_s, \mathbf{u}_s^{(\kappa-1)})$ based on \eqref{gradient_u} and
	set $\tau \gets \tau_{\mathrm{ini}}$;\\
	Design $\bar{\mathbf{u}}_s^{(\kappa-1)}$ by solving P3-C;\\
	Compute $\mathbf{u}_s^{(\kappa)} \gets \mathbf{u}_s^{(\kappa-1)} + \tau \left( \bar{\mathbf{u}}_s^{(\kappa-1)} - \mathbf{u}_s^{(\kappa-1)} \right)$;\\
	\hspace*{1em}{\bf while} $f(\mathbf{p}_s, \mathbf{u}_s^{(\kappa)}) - f(\mathbf{p}_s, \mathbf{u}_s^{(\kappa-1)}) < \hspace*{3em} \mu \tau \nabla_{\mathbf{u}_s} f(\mathbf{p}_s, \mathbf{u}_s^{(\kappa-1)})^{T} (\bar{\mathbf{u}}_s^{(\kappa-1)} - \mathbf{u}_s^{(\kappa-1)})$ {\bf do}\\
	\hspace*{1em} $\tau \gets \delta \tau$;\\
	\hspace*{1em}	Compute $\mathbf{u}_s^{(\kappa)} \gets \mathbf{u}_s^{(\kappa-1)} + \tau \left( \bar{\mathbf{u}}_s^{(\kappa-1)} - \mathbf{u}_s^{(\kappa-1)} \right)$;\\
	\hspace*{1em} {\bf end while}\\
	Update $\kappa \gets \kappa + 1$;\\
	{\bf end while}\\
	{\bf Return} $\mathbf{u}_s$.
\end{algorithm}

\begin{algorithm}[t]
	\caption{The Proposed AO Algorithm for Solving P0}
	{\bf Input:} $\{\mathbf{p}_s^{(0)}\}_{s \in \mathcal{S}}$, $\{\mathbf{u}_{s}^{(0)}\}_{s \in \mathcal{S}}$, $\mathbf{A}^{(0)}$, and the iteration numbers $I_o$ and $I_1$.\\ 
	{\bf while} $0 \leq i \textless I_o$ {\bf do}\\
	\hspace*{1em} Update $\mathbf{A}$ and $\mathbf{D}_{m}$ via Algorithm 1;\\
	\hspace*{1em} {\bf for} $s= 1:1:S$ {\bf do}\\
	\hspace*{3em} Given $\{\mathbf{p}_j\}_{j \in \mathcal{S}/s}$, $\{\mathbf{u}_{j}\}_{j \in \mathcal{S}}$, obtain $\mathbf{p}_s$ via \hspace*{3em} Algorithm 2;\\
	\hspace*{1em} {\bf end for}\\
	\hspace*{1em} {\bf for} $s= 1:1:S$ {\bf do}\\
	\hspace*{3em} Given $\{\mathbf{u}_j\}_{j \in \mathcal{S}/s}$, $\{\mathbf{p}_{j}\}_{j \in \mathcal{S}}$, obtain $\mathbf{u}_s$ via \hspace*{3em} Algorithm 3;\\
	\hspace*{1em} {\bf end for}\\
	{\bf end while}\\
	{\bf Return} $\mathbf{A}$, $\mathbf{D}_{m}$ $\mathbf{p}$, and $\mathbf{u}$.
\end{algorithm}

\subsection{Computational Complexity}
In this subsection, we analyze the computational complexity of the proposed algorithm,  which is primarily driven by SDR-based alternating minimization algorithm and feasible gradient descent method. Specifically, to obtain the digital beamforming vector $\mathbf{D}_{m}$, the computational complexity is $\mathcal{O}\left(I_1 ((N_{\rm RF} K)^{3.5} log(1/\varepsilon)+ N_t K)\right)$, where $\varepsilon$ and $I_{1}$ are the iteration accuracy and the number of iterations, respectively. The computational complexity is $\mathcal{O}\left(I_{2}  N_t K^2\right)$ for solving $\mathbf{p}$, where $I_{2}$ is the required number of iterations of the feasible gradient descent method. Similarly, the computational complexity is $\mathcal{O}\left(I_{3}  N_t K^2\right)$ for solving $\mathbf{u}$, where $I_{3}$ is the required number of iterations. Thus, the overall computational complexity of the proposed $\bf Algorithm \hspace*{0.02in} 1$ is $\mathcal{O}\left(I_o(I_1 ((N_{\rm RF} K)^{3.5} log(1/\varepsilon)+ N_t K)+(I_{2}+I_{3})N_t K^2) \right)$, where $I_o$ is the required iteration number of the outer iterations.

\section{Numerical Results}
\begin{figure}[htbp]
	\centering
	\includegraphics[height=2.2in,width=3.4in]{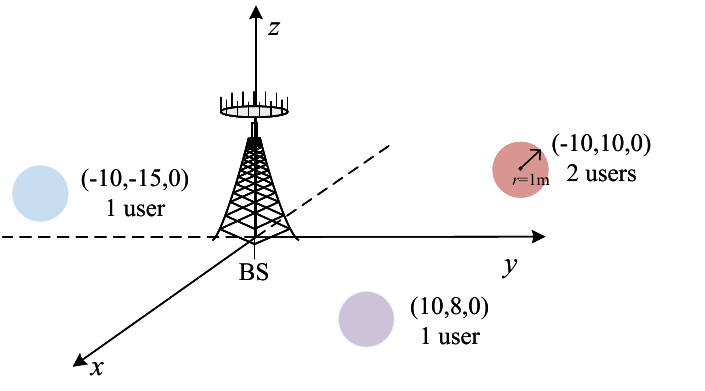}
	\caption{Simulation setup.}
	\label{fig_5}
\end{figure}

In this section, simulation results are provided to evaluate the performance of the proposed scheme. As presented in Fig. \ref{fig_5}, the BS is located at the origin. The users distribution is as follows: two users are randomly distributed within a sphere centered at $(-10, 10, 0)$, one user is randomly distributed within a sphere centered at $(10, 8, 0)$,  and one user is randomly distributed within a sphere centered at $(-10, -15, 0)$. The radius of all spheres is $r = 1$ m. The step size for the gradient approximation in \eqref{gradient} and \eqref{gradient_u} is set to $\epsilon=2^{-16}$. The default simulation parameters are listed in Table I unless particularly specified.

\begin{table}[htb]
	\begin{center}
		\caption{System parameters}
		\label{table:1}
		\begin{tabular}{|c|c|c|}
			\hline   \textbf{Parameters} & \textbf{Value} \\
			\hline  Each 6DMA surface antennas&  $N=16$ \\
			\hline   Number of 6DMA surfaces&  $S=4$ \\
			\hline   6DMA-BS site space&  Cube with 1m sides \\
			\hline   Central frequency & $f_{c}=300$ GHz  \\
			\hline   Bandwidth & $B=20$ GHz  \\
			\hline   Number of subcarriers & $M=8$  \\
			\hline   Number of users &  $K=4$ \\
			\hline   Number of RF chains & $N_{\rm{RF}}=4$  \\
			\hline   Maximum transmit power & $P_{\rm{max}}=$ 35 dBm  \\
			\hline   Noise power & $\sigma_{m,k}^{2} = -60$ dBm  \\
			\hline
		\end{tabular}
	\end{center}
\end{table}

\begin{figure}[t]
	\centering
	\includegraphics[height=2.5in,width=3.5in]{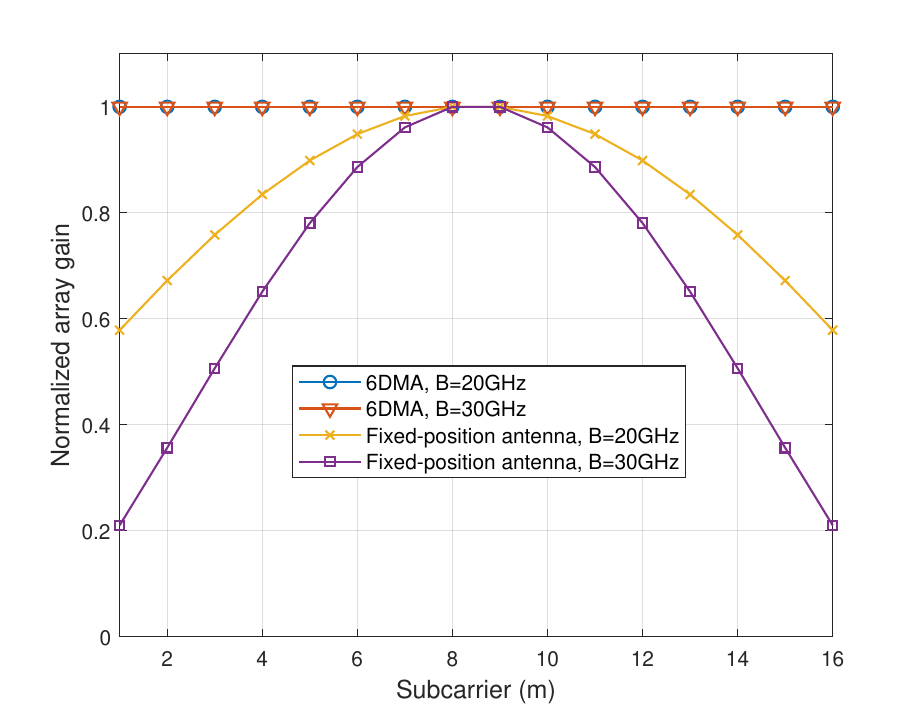}
	\caption{Normalized array gain versus subcarriers under different bandwidths.}
	\label{fig_6}
\end{figure}

Fig. \ref{fig_6} illustrates the normalized array gain versus subcarriers under different bandwidths. Here, we consider the single-user and single 6DMA surface scenario, i.e., $K = 1, S=1$, and the performance of conventional fixed-position antenna architecture is considered for comparison. It can be observed that the fixed-position antenna scheme suffers from the severe beam squint effects, leading to significant array gain degradation. This degradation becomes more pronounced when the bandwidth increases from $20$ GHz to $30$ GHz. In contrast, the 6DMA architecture effectively mitigates the beam squint effects, achieving nearly flat array gain across all subcarriers. The performance improvement arises from the flexible mobility and large movement range of 6DMA surfaces, which allow dynamic adjustment of surfaces positions to maintain beam alignment over the entire frequency band. These prove the superiority of 6DMA in wideband THz scenarios.
\begin{figure}[t]
	\centering
	\includegraphics[height=2.5in,width=3.5in]{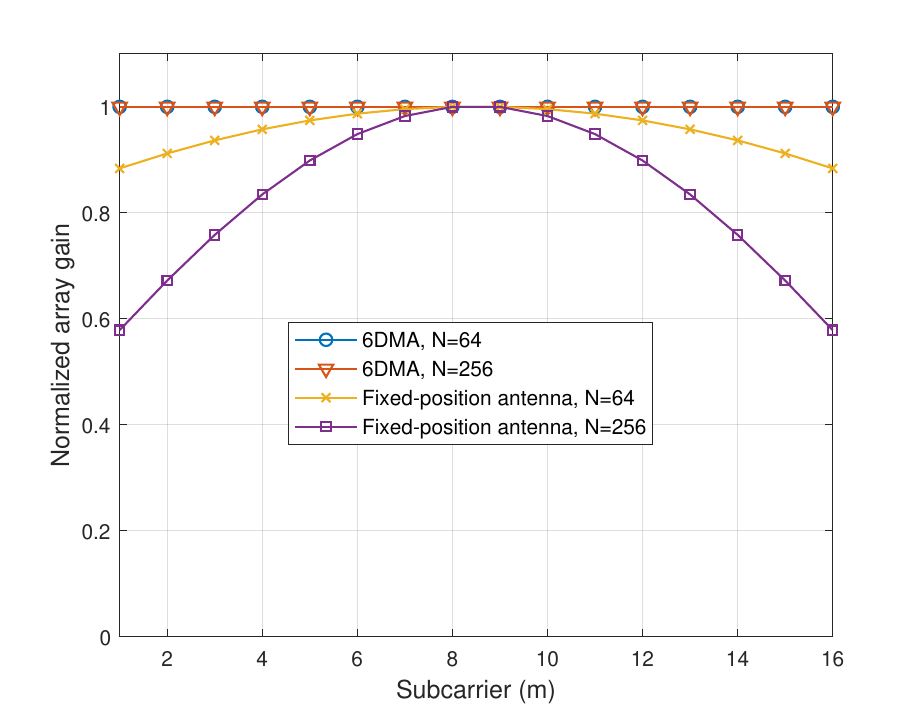}
	\caption{Normalized array gain versus subcarriers under different $N$.}
	\label{fig_7}
\end{figure}

Moreover, Fig. \ref{fig_7} presents the normalized array gain across subcarriers under different numbers of antennas. It can be found that the array gain degradation is more severe as the number of antennas grows under the conventional fixed-position antenna architecture. This phenomenon is mainly attributed to the beam squint effects, since a larger number of antenna elements results in greater delay spread across the array, thereby increasing the beam misalignment over frequency \cite{refI_8}. In contrast, the 6DMA architecture effectively suppresses this effects. Through dynamic adjustment of positions and rotations, 6DMA compensates for frequency-dependent phase delays, ensuring consistent beam alignment across all subcarriers. Consequently, the array gain achieved by 6DMA remain nearly flat even when the number of antennas increases from $64$ to $256$. It demonstrates the robustness of 6DMA against beam squint effects, highlighting its advantage over conventional fixed-position antenna systems in large-scale array deployments.

\begin{figure}[t]
	\centering
	\includegraphics[height=2.7in,width=3.6in]{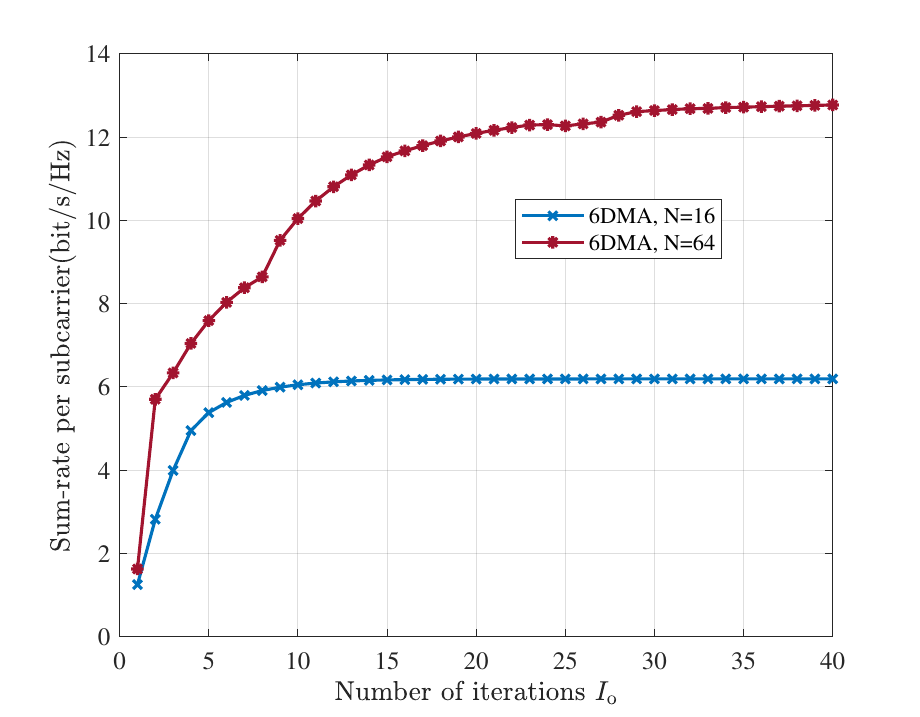}
	\caption{Sum rate versus the number of iterations.}
	\label{fig_8}
\end{figure}

Fig. \ref{fig_8} presents the convergence behavior of the proposed optimization algorithm (Algorithm 4) for different number of antennas $N$ deployed on each 6DMA surface, with the number of surfaces fixed at $S=4$. Here, we consider a multiuser scenario, i.e., $K = 4$. As shown in Fig. \ref{fig_8}, the sum-rate monotonically increases with the number of iterations and eventually converges to a stable value, thereby confirming the effectiveness of the proposed algorithm. Furthermore, a larger number of antennas on each  6DMA surface yields a higher sum-rate, which can be attributed to the increased spatial degrees of freedom and enhanced beamforming capability. It is shown that the proposed framework not only guarantees stable convergence but also effectively exploits larger antenna arrays to substantially improve system throughput.

\begin{figure}[t]
	\centering
	\includegraphics[height=2.7in,width=3.6in]{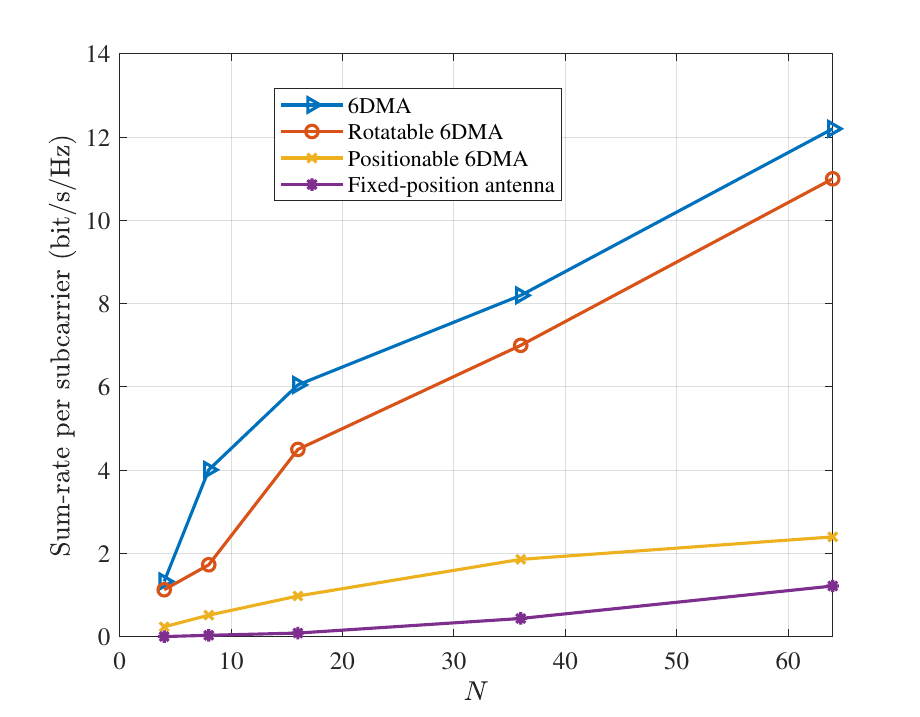}
	\caption{Sum rate versus the number of antennas of each 6DMA surface.}
	\label{fig_9}
\end{figure}
To further evaluate the effectiveness of the 6DMA-enabled multiuser downlink communications, we compare three representative benchmark schemes: 1) rotatable 6DMA with fixed positions, 2) positionable 6DMA with fixed rotations, and 3) conventional fixed-position antenna arrays with fixed position and rotation ($\beta=15^\circ$, $\alpha=0^\circ$). Fig. \ref{fig_9} depicts the sum-rate versus the number of antennas $N$ deployed on each 6DMA surface. The 6DMA architecture consistently outperform the conventional fixed-position antenna scheme. In particular, the positionable 6DMA provides notable improvements by mitigating beam misalignment through flexible displacement, while the rotatable 6DMA further enhances the performance by adaptively steering antenna rotations to optimize angular coverage. Moreover, the 6DMA architecture, which jointly exploits both positions and rotations adjustment, achieves the largest sum-rate. The above results indicate the advantages of incorporating six-dimensional movement into antenna design, as the additional spatial degrees of freedom effectively counteract the beam squint effects. 

\begin{figure}[t]
	\centering
	\includegraphics[height=2.7in,width=3.6in]{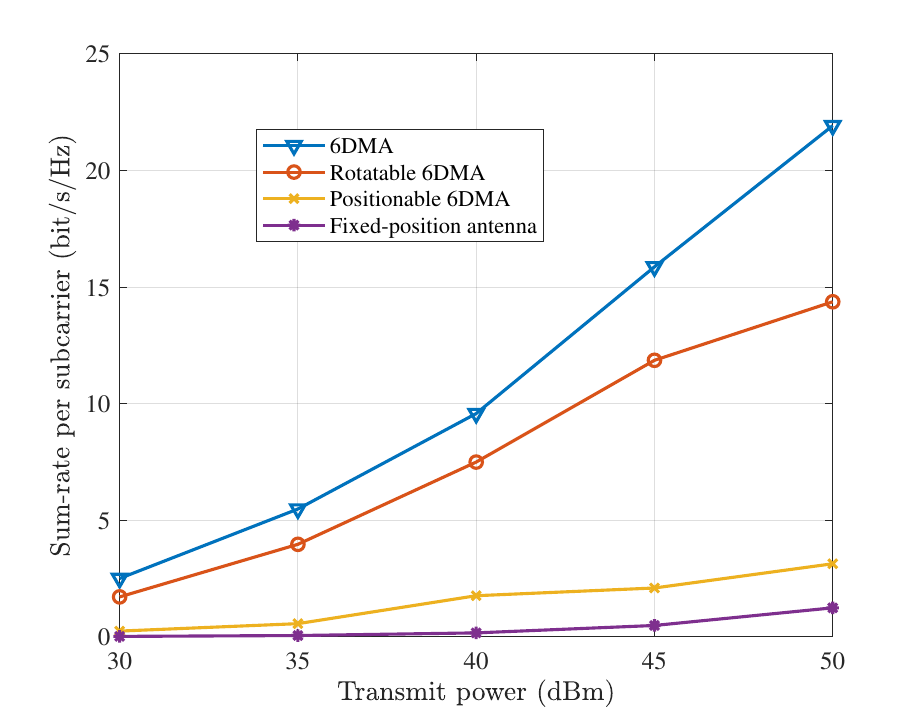}
	\caption{Sum rate versus the transmit power.}
	\label{fig_10}
\end{figure}

Fig. \ref{fig_10} shows the sum-rate against the transmit power. It can be noticed that  the sum-rate monotonically increases with the transmit power under all considered schemes. Besides, it can be found that 6DMA can achieve larger sum-rate as compared to the benchmark schemes under the same transmit power. Moreover, the performance gaps become more substantial as transmit power increases. This trend aligns with expectations that the sum-rate becomes more interference-limited as the transmit power increases, and adjusting the positions/rotations of 6DMA surfaces can significantly enhance the multiuser-MIMO channel condition at the BS for efficient interference suppression.

\section{Conclusions}
In this paper, we investigated the 6DMA-enabled wideband THz communication systems. We analyzed the normalized array gain characteristics of 6DMA, revealing that both positions and rotations significantly influence the beam squint effects. By flexibly adjusting 3D positions and 3D rotations of 6DMA surfaces, the wideband beam squint effects can be effectively mitigated. Moreover, we formulated a sum-rate maximization problem by jointly optimizing the hybrid beamforming and 3D positions and 3D rotations of the 6DMA surfaces. To address the inherent non-convexity of the problem, an AO framework was developed. Numerical results validated that the 6DMA system achieves superior array gain and sum-rate performance compared with conventional fixed-position antennas and partially movable 6DMA schemes, confirming its potential for practical THz communications. In future, we will focus on optimizing the antenna positions and rotations in the 6DMA THz system through
hybrid-field 6DMA channel model.

\end{document}